\documentclass[twocolumn,superscriptaddress,showpacs,floatfix,aps,prb]{revtex4}
\usepackage{graphicx}
\usepackage{color}
\usepackage{siunitx}
\DeclareSIUnit\muB{$\mu_B$}
\usepackage{dcolumn}
\usepackage{bm}
\usepackage{amsmath}
\usepackage{amsfonts}
\usepackage{amssymb}
\usepackage[normalem]{ulem}
\usepackage[colorlinks=true]{hyperref}
\usepackage{url}
\newcommand{\etal}{\textit{et al.}}
\newcommand{\Pcoldown}{\multicolumn{1}{c}{$P_\downarrow$}}
\newcommand{\Pcolup}{\multicolumn{1}{c}{$P_\uparrow$}}
\renewcommand{\vec}[1]{{\bm{#1}}}
\newcommand{\MAE}{E_\mathrm{MAE}}
\begin{document}
\title{Study of electronic and magnetic properties and related x-ray absorption spectroscopy
of ultrathin Co films on BaTiO$_3$}
\author{M.~Hoffmann}
\affiliation{Institut f\"ur Physik, Martin-Luther-Universit\"at Halle-Wittenberg, D-06099 Halle, Germany}
\affiliation{ Max-Planck Institut f\"ur Mikrostrukturphysik, Weinberg 2, D-06120 Halle}
\author{St.~Borek}
\affiliation{Department Chemie, Ludwig-Maximilians-Universit\"at M\"unchen, Butenandtstra\ss{}e 11, D-81377 M\"unchen, Germany}
\author{I.\,V.~Maznichenko}
\affiliation{Institut f\"ur Physik, Martin-Luther-Universit\"at Halle-Wittenberg, D-06099 Halle, Germany}
\author{S.~Ostanin}
\affiliation{ Max-Planck Institut f\"ur Mikrostrukturphysik, Weinberg 2, D-06120 Halle}
\author{G.~Fischer}
\affiliation{Institut National des Sciences Appliqu\'ees Toulouse, LPCNO, 135 Avenue de Rangueil, 31077 Toulouse, France}
\author{M.~Geilhufe}
\affiliation{ Max-Planck Institut f\"ur Mikrostrukturphysik, Weinberg 2, D-06120 Halle}
\author{W.~Hergert}
\affiliation{Institut f\"ur Physik, Martin-Luther-Universit\"at Halle-Wittenberg, D-06099 Halle, Germany}
\author{I.~Mertig}
\affiliation{Institut f\"ur Physik, Martin-Luther-Universit\"at Halle-Wittenberg, D-06099 Halle, Germany}
\affiliation{ Max-Planck Institut f\"ur Mikrostrukturphysik, Weinberg 2, D-06120 Halle}
\author{A.~Ernst}
\affiliation{ Max-Planck Institut f\"ur Mikrostrukturphysik, Weinberg 2, D-06120 Halle}
\author{A.~Chass{\'e}}
\affiliation{Institut f\"ur Physik, Martin-Luther-Universit\"at Halle-Wittenberg, D-06099 Halle, Germany}
\email{angelika.chasse@physik.uni-halle.de}
\date{\today}
\begin{abstract}
  We present a first-principles study of electronic and magnetic
  properties of thin Co films on a BaTiO$_3$(001) single crystal.  The
  crystalline structure of 1, 2, and 3 monolayer thick Co films was
  determined and served as input for calculations of the electronic
  and magnetic properties of the films. The estimation of exchange
  constants indicates that the Co films are ferromagnetic with a high
  critical temperature, which depends on the film thickness and the
  interface geometry.  In addition, we calculated x-ray absorption
  spectra, related magnetic circular dichroism (XMCD) and linear
  dichroism (XLD) of the Co L$_{2,3}$ edges as a function of Co film
  thickness and ferroelectric polarization of BaTiO$_3$.  We found
  characteristic features, which depend strongly on the magnetic
  properties and the structure of the film. While there is only a weak
  dependence of XMCD spectra on the ferroelectric polarization, the
  XLD of the films is much more sensitive to the polarization
  switching, which could possibly be observed experimentally.
\end{abstract}
\pacs{75.50.Cc, 71.20.Lp, 71.15.Rf}
\maketitle

The interface between a magnetic thin film and a ferroelectric
material is the subject of several recent
investigations.~\cite{Tsymbal2006,Duan2006,Duan2006a,Sahoo2007,Velev2008,
  Fechner2008,Fechner2009,Gruverman2009,Niranjan2009,
  Garcia2010,Vaz2010,Fechner2010,Cao2011,Meyerheim2011,Valencia2011,
  Borek2012,Lukashev2012,Dai2012,Lu2012} The major part of these
studies is devoted to Fe/ferroelectric
interfaces,~\cite{Duan2006,Sahoo2007,Velev2008,Fechner2008,Fechner2009,
  Garcia2010,Fechner2010,Meyerheim2011,Valencia2011,Borek2012,Lukashev2012,Dai2012}
since ferromagnetic iron is supposed to be a good candidate as a
ferromagnetic electrode in two-component multiferroics. Although it
was shown, that a ferroelectric film can be grown on an iron substrate
and a functional heterojunction can be
fabricated,~\cite{Meyerheim2011,Valencia2011} the multiferroic effects
by polarization switching were found to be not strong enough, since
the main changes of the functional properties occur only in the
vicinity of the interface.~\cite{Duan2006,Fechner2008,Valencia2011}
Another impediment is non-ideal compatibility between an iron and an
oxide surface.  Until now it was not shown that junctions with
symmetric interfaces, a highly desirable condition for coherent
electronic transport,~\cite{Tusche2005} can be grown.  Accordingly,
ferromagnetic oxides La$_{1-x}$Sr$_x$MnO$_3$ and SrRuO$_3$ were used
as electrodes in multiferroic
junctions.~\cite{Duan2006a,Gruverman2009,Niranjan2009,Vaz2010,Lu2012}
Such interfaces can be almost ideally grown, but the Curie temperature
of these oxides is too low for functional devices. Therefore, search
for an appropriate ferromagnetic/ferroelectric interface is still in
progress.

Among the above cited studies only few investigations deal with
ultrathin ferromagnetic films on a ferroelectric
substrate.~\cite{Fechner2008,Fechner2009,Fechner2010,Borek2012} In
particular, it was shown that two monolayer thick Fe films on the
BaTiO$_3$(001) surface might not be ferromagnetic because of the film
geometry and magneto-elastic properties of iron.~\cite{Fechner2008} Additionally, a
substantial charge and spin moment transfer was found at the interface
by altering the polarization direction.~\cite{Fechner2008} 

Thus, ultrathin films of Fe on a BaTiO$_3$(001) single crystal are
magnetically unstable,~\cite{Fechner2008} but cobalt exhibits usually
stable ferromagnetic characteristics in many nanostructures.
Therefore, we continue our work on ultrathin metallic films on
ferroelectric single crystals and suggest in this paper to use Co as
the ferromagnetic material on BaTiO$_3$(001).

Despite the comprehensive review about the progress in this field
given by Vaz \etal~\cite{Vaz2008} there are only few information about
the interface of Co and perovskites like BaTiO$_3$ (BTO).  In the
framework of spin-polarized DFT calculations, as implemented in the
Vienna {\it ab initio} Simulation Package (VASP), multiferroic tunnel
junctions of Co/BTO/Co were investigated by Cao \etal~\cite{Cao2011}
They showed that a critical thickness of BTO unit cells is necessary
for the appearance of ferroelectricity, which is inhibited by a
depolarizing electrostatic field, caused by dipoles at the
ferroelectric-metal interfaces.~\cite{Junquera2003}
In the work of Oleinik \etal~\cite{Oleinik2001} first-principles
calculations were applied on Co/STO/Co(001) magnetic tunnel junctions,
where a strong covalent bond between Co and O 
and an induced magnetic moment of $|m_\mathrm{s}|=\SI{0.25}{\muB}$
at the Ti atom was observed. This is similar to the case of Fe/BTO/Fe
  tunnel junctions. In another work, Polisetty
\etal~\cite{Polisetty2010} applied piezoelectrically controlled strain
for electric tuning of exchange-bias fields of BTO/Co/CoO
heterostructures.

In our first-principles study, we investigated systematically the
crystalline structure of ultrathin Co films on BTO and their
electronic and magnetic properties in dependence on the film thickness
and the polarization of BTO.  The calculations were performed using a
so-called multi-code approach, in which atomic positions were obtained
using a pseudo-potential
code, VASP.\cite{Kresse1994,Kresse1996} 
This information serves as input for electronic and magnetic
structure calculations with different multiple-scattering Green
function methods.~\cite{Luders2001,Ebert2011,Ebert-kkr}

In addition, we simulated x-ray absorption spectra~(XAS) and the
related x-ray magnetic circular dichroism~(XMCD), which is the method
of choice to prove experimentally the change of the magnetic
structure. The x-ray linear dichroism (XLD) is dicussed in respect to
the occupation of $d$ orbitals.  These methods are local and
site-sensitive and offer the opportunity to investigate both the
magnetic and structural properties as a function of layer
thickness.~\cite{Borek2012} So, we traced the dependence of XAS of Co
L$_{2,3}$ edges as a function of the Co film thickness and the
polarization direction of BTO. All results were compared with the
previous results for Fe thin films on BTO.~\cite{Fechner2008,Borek2012}

Our paper is organized as follows. Our multi-code approach and
computational details are presented in the next section~\ref{methods}.
Then, in sections~\ref{struc} and \ref{results}, we discuss
structural, electronic, and magnetic properties of ultrathin Co films
on BTO.  Sections~\ref{xmcd} and \ref{xmld} deal with computational
simulations of XAS, XMCD, and XLD spectra.  Conclusion is offered in
section~\ref{concl}.

\section{Computational details}
\label{methods}

\subsection{Structural optimization}
The information about the crystalline structure of Co/BTO(001) was
obtained using projector-augmented wave
pseudo-potentials~\cite{Blochl1994} implemented in the VASP
code.~\cite{Kresse1994,Kresse1996} The plane-wave basis was taken with
a cutoff energy of \SI{400}{\electronvolt}.  The calculations were
performed within the local spin-density approximation (LSDA). We used
the parametrization of Perdew and Wang for the exchange-correlation
functional (PW-functional).~\cite{Perdew1992} Here, we believe that
the impact of electronic correlations on the interface magnetoelectric
coupling is minor.

We performed the structural optimization of Co/BTO(001) in dependence
on the Co film thickness and the ferroelectric polarization $\vec{P}$
of BTO(001).  To model the (001) surface of polar BTO, we constructed
a 5 unit cells ($\sim\SI{2}{\nano\meter}$) thick supercell, with a
vacuum spacer of \SI{2}{\nano\meter} along the [001] direction.  The
lattice parameters were set to the theoretical equilibrium values of
tetragonal BTO:~\cite{Fechner2008} $a=\SI{3.943}{\angstrom}$ and
$c/a=1.013$.  The intralayer cation-oxygen displacements
$\delta = z_\mathrm{O} -z_\mathrm{cation}$ (see Fig.~\ref{fig_str}) in BTO
cause the ferroelectric polarization, $\vec{P}$, along [001]. When
$\vec{P}$ is antiparallel to the surface normal ($P_\downarrow$), the
oxygen in each monolayer (ML) are higher then the cations
($\delta>0$), and {\it vice versa}, the ferroelectric state
$P_\uparrow$ means $\delta <0$.  Since the (001) surface of BTO is
TiO$_2$ terminated~\cite{Fechner2008} for both directions of
$\vec{P}$, the Co atoms find their relaxed positions atop the oxygen
atoms.  This is in agreement with the recent experimental data for
Fe/BTO(001).~\cite{Meyerheim2011} Each Co monolayer of the slab
contains, therefore, two atoms per unit cell while the Co film
thickness, $L$, varies between one and three ML.  For Co/BTO(001) the
atomic positions of the four top BTO layers and all Co ML were allowed
to relax.  After the relaxation, the calculated forces are always less
than \SI[per-mode=symbol]{0.5d-2}{\electronvolt\per\angstrom}.  The
Brillouin zone of the slab was sampled with a 10$\times$10$\times$6
$k$-point Monkhorst-Pack mesh~\cite{Monkhorst1976} during the force
minimization.

\subsection{Electronic and magnetic structure calculations}
The relaxed geometry was used for further first-principles
calculations using the multiple-scattering Green function method
(Hutsepot) within the atomic sphere approximation to the crystal
potential.~\cite{Korringa1947,Kohn1954,Luders2001} We took an angular
momentum cutoff of $l_\mathrm{max} = 3$ for the Green function, a {\it
  k}-point mesh of $24\times 24\times 12$ for the BZ integration and
24 Gaussian quadrature points for complex energy contour integration.
With the Green function $G(E)$ of the system, all quantities of
interest follow in a straightforward way.  In particular, we are
interested in ground state properties like the density of states (DOS)
and the local magnetic moments~\cite{Guo1998} in dependence on Co film
thickness $L$ and ferroelectric polarization direction $\vec{P}$ of
BTO.

\begin{figure}
 \center{\includegraphics[width=0.950\columnwidth]{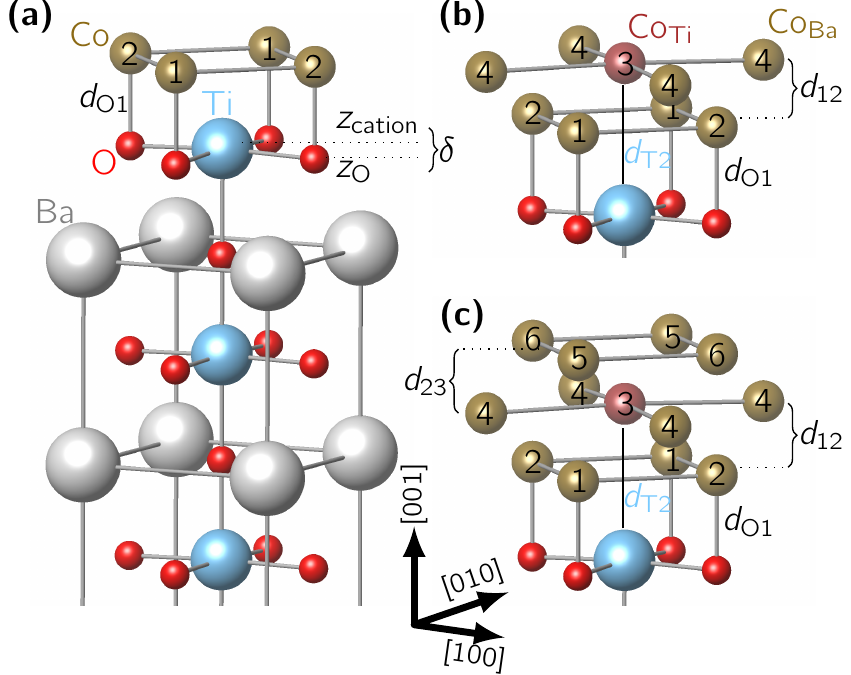}}
 \caption{Geometry of Co films on the BaTiO$_3$ surface for 1~ML (a),
   2~ML (b), and 3~ML (c) cobalt thicknesses, respectively. Spheres
   represent Co (brown), Ba (gray), Ti (blue), and oxygen (red)
   sites. Values of selected interatomic distances for polarizations $\vec{P} =
   P_{\downarrow}, P_{\uparrow}$ are presented in Table~\ref{tab_d}.
   The picture of the underlying structure was done with 
   VESTA.~\cite{Momma2011} The numbers for the Co atoms 
   symbolize the atoms of the unit cell and correspond to those in
   the figures of the supplementary material.}
\label{fig_str}
\end{figure}

To describe magnetic properties of Co/BTO(001), we calculated
interatomic exchange coupling parameters $J_{ij}$ using the magnetic
force theorem implemented within the Green function
method.~\cite{Liechtenstein1987} The exchange coupling constants
$J_{ij}$ can be used to obtain spin-wave spectra by the
diagonalization of the Heisenberg Hamiltonian
\begin{eqnarray} 
  H = -{\sum_{i\neq j}} J_{ij}{\vec{e}_i}{\vec{e}_j} - \sum_i \MAE \,(e_i^z)^2 \, ,  
  \label{eq2}
\end{eqnarray}
where $i$ and $j$ label magnetic atoms, ${\vec{e}_i}$ is a unit vector
in the direction of the magnetic moment of the $i$-th atom and
$\MAE$ is the magnetic anisotropy energy (MAE), which is positive
(negative) for the case of the easy-axis (easy-plane) anisotropy type.
It can be calculated directly with the KKR method by using again the
magnetic force theorem.~\cite{Razee1997} The critical temperatures
were estimated using a Monte Carlo~(MC) simulation with the model
Hamiltonian in Eq. \eqref{eq2}.  For the simulation, two dimensional
supercells, which repeat the unit cell (see Fig.~\ref{fig_str})
$60\times60$, $80\times80$, and $100\times100$ times in $x$ and $y$
direction were constructed.  In those directions, we considered also
periodic boundary conditions and restricted the calculation only to
the magnetic atoms of the unit cell (2 Co atoms for $L=1$, 4 Co atoms
for $L=2$, and 6 Co atoms for $L=3$).  Within the supercell, the
magnetic moment at lattice site $i$ interacts with its neighbors at
the site $j$ via the first-principles $J_{ij}$. During a MC run, a
lattice site $j$ with the magnetic moment vector $\vec{e}_j$ was
chosen and a new random direction $\vec{e}'_j$ was created. The energy
of the system determines whether $\vec{e}'_j$ or $\vec{e}_j$ was
kept. Performing this procedure $N$ times on a lattice of $N$ sites is
defined as one MC step. The starting point was a high-temperature
disordered state above the critical temperature $T_\mathrm{C}$.  In the
course of the simulations, the temperature was stepwise reduced until
magnetic ordering was reached. For each temperature $T$, the thermal
equilibrium was assumed to be reached after \SI{20000}{} MC
steps. The thermal averages were determined over \SI{20000}{}
additional MC steps.  $T_\mathrm{C}$ was then obtained from the
temperature dependency of the magnetic susceptibility.  With respect
to fitting procedures and finite temperature sampling, all three
supercells yielded the same critical temperature within an uncertainty
range of $\pm\SI{5}{\kelvin}$.  More details of our MC scheme can be
found in Refs.~\onlinecite{Fischer2009,Otrokov2011,Otrokov2011a}.

\subsection{X-ray absorption spectra simulations}
Using the self-consistent potentials of the scalar-relativistic Green
function method (Hutsepot), the calculations were extended to a spin-resolved
fully relativistic KKR code~(SPR-KKR).~\cite{Ebert-kkr}
This Green function method~\cite{Ebert-kkr} allows the calculation 
of x-ray
absorption coefficients $\mu^\lambda(E)$ in dependence on the energy $E$
and the polarization $\lambda$ of the x-rays.~\cite{Koningsberger1988,Stoehr1998,Laan1996}
For SPR-KKR calculations of magnetic moments, XAS and DOS, we have taken
144, 578 and 225~$k$-points in the irreducible part of the Brillouin
zone, respectively. The calculation of the structure constants were done using the
Ewald method with parameter 2.0 and 4.0 in real space and reciprocal
space, respectively.  For the Ewald parameter connecting the summation
in real and reciprocal space a value of 0.8 was used, making the
summation in the real space converge faster. These parameters were
applied for all considered Co/BTO systems.

\section{Crystalline structure of ultrathin Co films on BTO}
\label{struc}

\begin{table}
  \caption{Relaxed interatomic distances (in \si{\angstrom}) of  
  (Co)$_{L}$/BTO(001) ($L\le3$ and $\vec{P} = P_{\downarrow}, P_{\uparrow}$). 
  For each $L$, the distances between Co ML $i$ and $j$ are denoted by $d_{ij}$ while
  the $z$-separation from interfacial O (Ti) to Co of the first (second) ML is shown by
  $d_{\mathrm{O}1}$ ($d_{\mathrm{T}2}$). For comparison, 
  also nearest neighbor Co distances in bulk (hcp)
  from Ref. \onlinecite{Lejaeghere2014} and distances for (Fe)$_{L}$/BTO(001)
  are shown. }
  \label{tab_d}
  \begin{ruledtabular}
    \begin{tabular*}{\columnwidth}{@{\extracolsep{\fill}}lccccc}
	  &                   &\multicolumn{2}{c}{Co}     &\multicolumn{2}{c}{Fe}     \\
	  &                   &$P_\downarrow$&$P_\uparrow$&$P_\downarrow$&$P_\uparrow$\\
      \hline
      $L=1$ &$d_{\mathrm{O}1}$&1.778&1.784&1.774&1.781\\
      \hline
      $L=2$ &$d_{\mathrm{O}1}$&1.856&1.853&1.857&1.855\\
	    &$d_{12}$         &1.116&1.107&1.049&1.054\\
	    &$d_{\mathrm{T}2}$&3.022&2.927&2.971&2.918\\
      \hline
      $L=3$ &$d_{\mathrm{O}1}$&1.827&1.829&1.843&1.849\\
	    &$d_{12}$         &1.183&1.179&1.218&1.241\\
	    &$d_{23}$         &1.154&1.153&1.134&1.114\\
	    &$d_{\mathrm{T}2}$&3.094&3.014&3.214&3.121\\
      \hline
      \multicolumn{2}{l}{bulk (hcp):} & \multicolumn{2}{c}{2.478} & & \\
    \end{tabular*}
  \end{ruledtabular}
\end{table} 

For the ultrathin Co films deposited on BTO(001), the optimized
interatomic distances are collected in Tab.~\ref{tab_d}, in comparison
with the corresponding structure of
Fe/BTO(001).\cite{Fechner2008,Borek2012} The definition of our
distances are given in Fig.~\ref{fig_str}.  Below in the text, the
interface TiO$_2$ layer is denoted as I for all films.  The Co layers
towards the surface are labeled $\mathrm{I}+1$, $\mathrm{I}+2$, and $\mathrm{I}+3$.

Structural relaxation, with the in-plane degrees of freedom for the
layers starting with $I-2$, energetically favors the structure with
the Co sites in layer $\mathrm{I}+1$ on top of the O sites in layer I.
In the case of 1~ML Co ($L=1$), the structure is similar to that of
Fe/BTO(001).\cite{Fechner2008,Borek2012} For both multiferroic
systems, the polarization state $P_\downarrow$ results in a slightly
shorter distance, $d_{\mathrm{O}1}$, between Co (Fe) and interfacial
O.

When the Co film thickness increases to $L=2$ and 3, a distorted
bodycentered tetragonal (bct) lattice is formed, although it is not
typical for Co.~\cite{Vaz2008} We found that the polarization reversal
affects mainly the positions of interfacial Ti (I) and consequently
those of the Co~$(\mathrm{I}+2)$ atoms, labeled as Co$_\mathrm{Ba}$ and
Co$_\mathrm{Ti}$, respectively (see Fig.~\ref{fig_str}).  This is
important concerning the magnetic properties of the films.
\section{Electronic and magnetic properties of ultrathin Co films on BTO}
\label{results}

In order to determine the magnetic ground state for the three film
thicknesses, we calculated the total energy for the ferromagnetic (FM)
and antiferromagnetic (AFM) configurations of the magnetic moments of
the Co atoms.  In all cases, the FM solution was preferable,
independent of thickness or polarization.  So, the thin cobalt films
are strong ferromagnets and show a lower sensitivity to structural
transformations then the iron layers, which showed a considerable
change in the magnetic order for different layer thicknesses.  For
the 2~ML Fe film on BTO, the $m_\mathrm{Fe}$ in layer $\mathrm{I}+1$ was
almost quenched while the sizable moments in the surface layer $\mathrm{I}+2$
are ordered antiparallel. This results in a total magnetic moment of
$m\rightarrow 0$.  Deposition of a third Fe monolayer restored the
ferromagnetic order of the 1~ML Fe film.

\subsection{Magnetic moments and density of states}

\newcolumntype{C}[1]{>{\arraybackslash}m{#1}}

\begin{table}
  \caption{SPR-KKR results of magnetic spin moments $m_\mathrm{s}$ ($\si{\muB}/\mathrm{atom}$) for (Co)$_L$/BTO(001) 
    in dependence on the Co film thickness $L$ and the direction of ferroelectric polarization $\vec{P}$ ($\vec{M}\parallel[001]$).
    The Co labels follow from Fig.~\ref{fig_str}. In $\mathrm{I}+1$ and $\mathrm{I}+3$ the Co atoms are equivalent.}
  \label{tab_ms}
  \begin{ruledtabular}
    \begin{tabular*}{\columnwidth}{@{\extracolsep{\fill}}l l *{6}{D{.}{.}{-1}} }
      && \multicolumn{2}{c}{$L=1$} & \multicolumn{2}{c}{$L=2$} & \multicolumn{2}{c}{$L=3$} \\
      && \Pcoldown & \Pcolup & \Pcoldown & \Pcolup & \Pcoldown & \Pcolup \\ \hline
      Co& $\mathrm{I}+3$              &       &        &       &       &  1.39 &  1.42\\ 
      Co$_\mathrm{Ba}$& $\mathrm{I}+2$&       &        &  2.00 &  2.03 &  1.60 &  1.64\\ 
      Co$_\mathrm{Ti}$& $\mathrm{I}+2$&       &        &  1.89 &  1.93 &  1.45 &  1.47\\ 
      Co& $\mathrm{I}+1$              &  1.71 &  1.71  &  1.62 &  1.62 &  1.42 &  1.37\\ 
      O & I                         &  0.08 &  0.08  &  0.03 &  0.03 &  0.06 &  0.05\\
      Ti& I                         & -0.11 & -0.18  & -0.12 & -0.16 & -0.03 & -0.08\\ \hline
      \multicolumn{2}{@{\extracolsep{\fill}}l}{Co bulk (hcp):}      &  1.62 &&&&&
    \end{tabular*}
  \end{ruledtabular}
\end{table}

\begin{table}
  \caption{SPR-KKR results of magnetic orbital moments $m_\mathrm{o}$ ($\si{\muB}/\mathrm{atom}$) for (Co)$_L$/BTO(001) 
    in dependence on the Co film thickness $L$ and the direction of ferroelectric polarization $\vec{P}$
    ($\vec{M}\parallel[001]$).
The Co labels follow from Fig.~\ref{fig_str}. In $\mathrm{I}+1$ and $\mathrm{I}+3$ the Co atoms are equivalent.}
  \label{tab_mo}
  \begin{ruledtabular}
    \begin{tabular*}{\columnwidth}{@{\extracolsep{\fill}}l l *{6}{D{.}{.}{-1}} } 
      && \multicolumn{2}{c}{$L=1$} & \multicolumn{2}{c}{$L=2$} & \multicolumn{2}{c}{$L=3$} \\
      && \Pcoldown & \Pcolup & \Pcoldown & \Pcolup & \Pcoldown & \Pcolup \\ \hline
      Co& $\mathrm{I}+3$               &      &      &      &      & 0.09 & 0.08\\ 
      Co$_\mathrm{Ba}$& $\mathrm{I}+2$ &      &      & 0.08 & 0.08 & 0.07 & 0.07\\ 
      Co$_\mathrm{Ti}$& $\mathrm{I}+2$ &      &      & 0.09 & 0.08 & 0.07 & 0.07\\ 
      Co& $\mathrm{I}+1$               & 0.17 & 0.17 & 0.07 & 0.06 & 0.09 & 0.09\\ \hline
      \multicolumn{2}{@{\extracolsep{\fill}}l}{Co bulk (hcp):}          & 0.06 &&&&& \\ 
    \end{tabular*}
  \end{ruledtabular}
\end{table}

\begin{table}
  \caption{SPR-KKR results of magnetic orbital moments $m_\mathrm{o}$ ($\si{\muB}/\mathrm{atom}$) 
  with $\vec{M}\parallel[100]$. 
  The Co labels follow from Fig.~\ref{fig_str}. In this field direction the Co atoms are nonequivalent.}
  \label{tab_mo1}
  \begin{ruledtabular}
    \begin{tabular*}{\columnwidth}{@{\extracolsep{\fill}}l l *{6}{D{.}{.}{-1}} } 
      && \multicolumn{2}{c}{$L=1$} & \multicolumn{2}{c}{$L=2$} & \multicolumn{2}{c}{$L=3$} \\
      && \Pcoldown & \Pcolup & \Pcoldown & \Pcolup & \Pcoldown & \Pcolup \\ \hline
      Co$_6$& $\mathrm{I}+3$               &      &      &      &      & 0.12 & 0.07\\ 
      Co$_5$& $\mathrm{I}+3$               &      &      &      &      & 0.07 & 0.11\\ 
      Co$_4$& $\mathrm{I}+2$ &      &      & 0.12 & 0.12 & 0.07 & 0.07\\ 
      Co$_3$& $\mathrm{I}+2$ &      &      & 0.12 & 0.13 & 0.06 & 0.06\\ 
      Co$_2$& $\mathrm{I}+1$               & 0.16 & 0.12 & 0.08 & 0.08 & 0.06 & 0.04\\ 
      Co$_1$& $\mathrm{I}+1$               & 0.13 & 0.17 & 0.07 & 0.06 & 0.07 & 0.05\\
    \end{tabular*}
  \end{ruledtabular}
\end{table}

The calculated spin moments (see Tab.~\ref{tab_ms}) for the Co atoms
show a strong dependence on the geometry of the films and the
hybridization of Co $3d$ states with the electronic bands of the
substrate.  The different environment of e.g. Co$_\mathrm{Ba}$ and
Co$_\mathrm{Ti}$, in particular, the atomic volumes and the
band hybridization, influences also these moments.

In the case of $L=1$ and $L=2$\,, the spin moments for Co atoms at the
surface, are larger then for the Co bulk (hcp) because of the symmetry
reduction and an enhanced volume per Co atom.  The Co magnetic moments
in layer $\mathrm{I}+1$ for $L=2$ approach their bulk value.  For $L=3$,
the spin moments of Co~$(\mathrm{I}+1)$ and Co$_\mathrm{Ti}$ are smaller
in comparison to $L=2$, because of slightly elongated distances
$d_{12}$ and $d_{\mathrm{T}2}$, respectively.  In the case of $L=3$\,,
we observe for all three layers a quenching of the spin moment in
comparison to Co bulk (hcp), which results from a strong reduction of
the volume per Co atom.

\begin{figure}
  \center{\includegraphics[width=0.950\columnwidth]{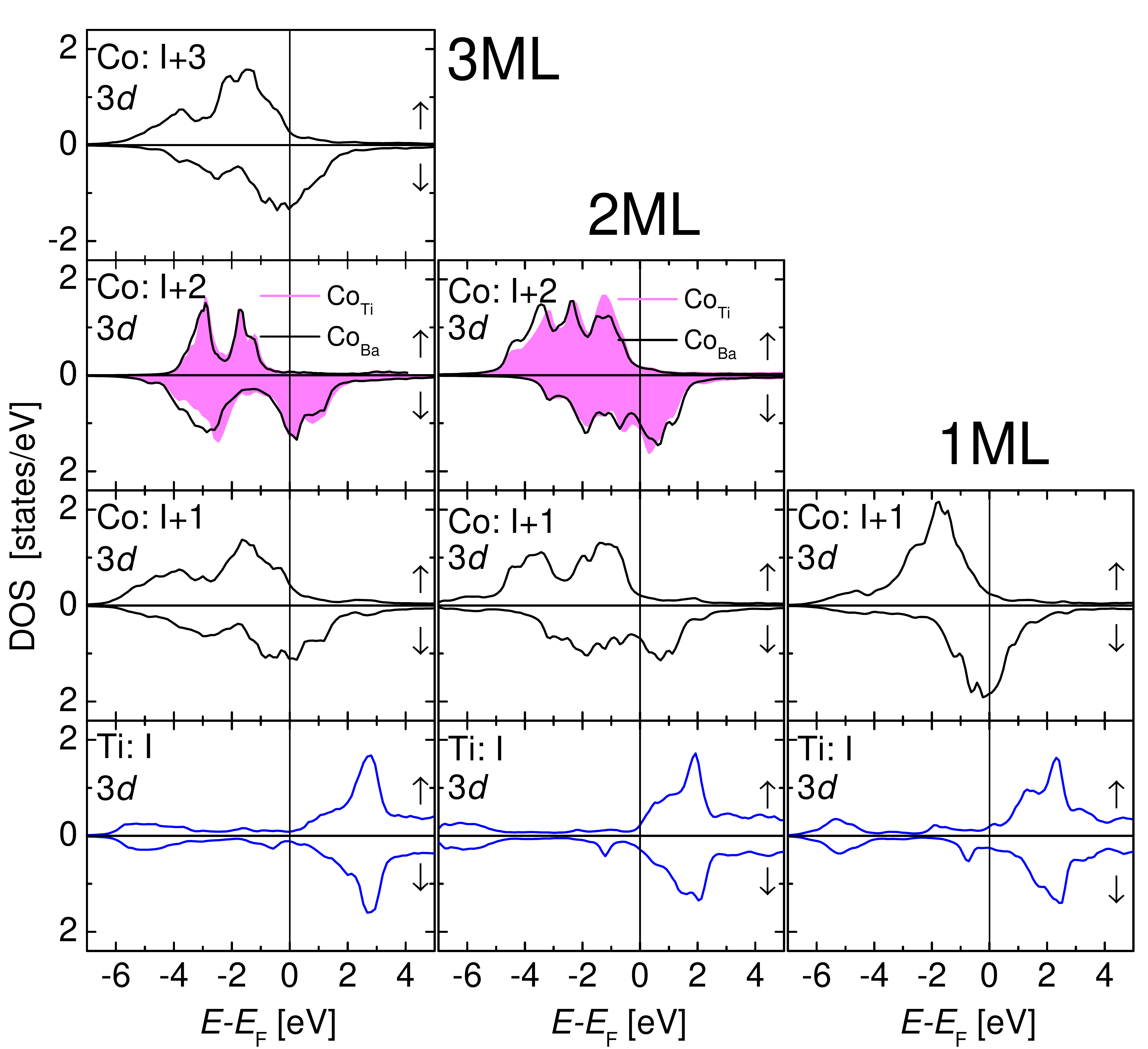}}
  \caption{Spin-resolved ($\uparrow$, $\downarrow$) density of
    $d$ states for Co and Ti atoms in dependence on film thickness
    $L$. It is only shown for $P_\uparrow$, because there are
      only small differences for $P_\downarrow$.}
  \label{fig_dos3d}
\end{figure}

\begin{figure}
 \center{\includegraphics[width=0.7\columnwidth]{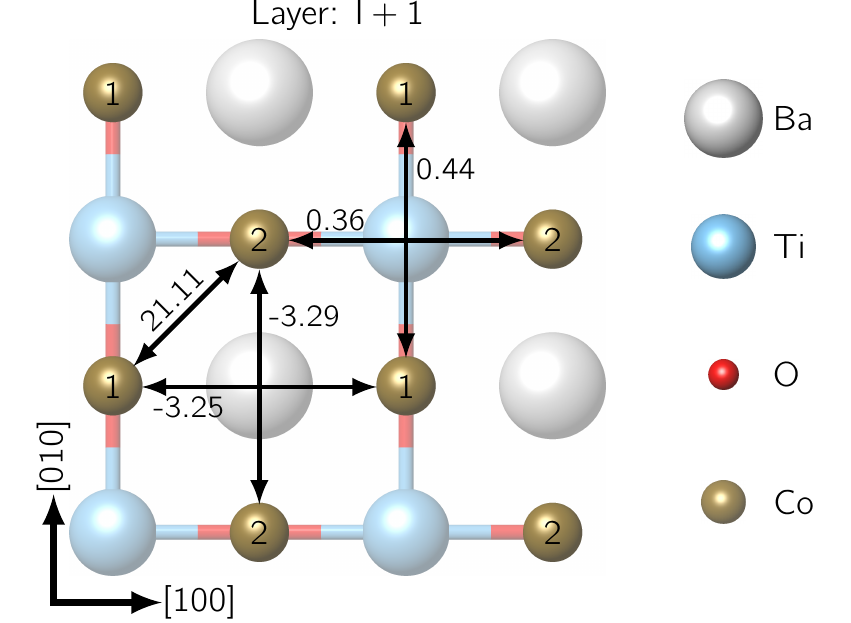}}
 \caption{Calculated intralayer exchange constants (in
   \si{\milli\electronvolt}) for the $\mathrm{I}+1$ layer in 1~ML Co/BTO(001)  and $P_\uparrow$, 
   viewed from the $z$ direction.
   Atoms in lighter colors are below those in darker colors. 
   The numbers of the Co atoms 
   correspond to those used in Fig.~\ref{fig_str}(a).}
\label{fig:J1}
\end{figure}

\begin{figure}
 \center{\includegraphics[width=0.7\columnwidth]{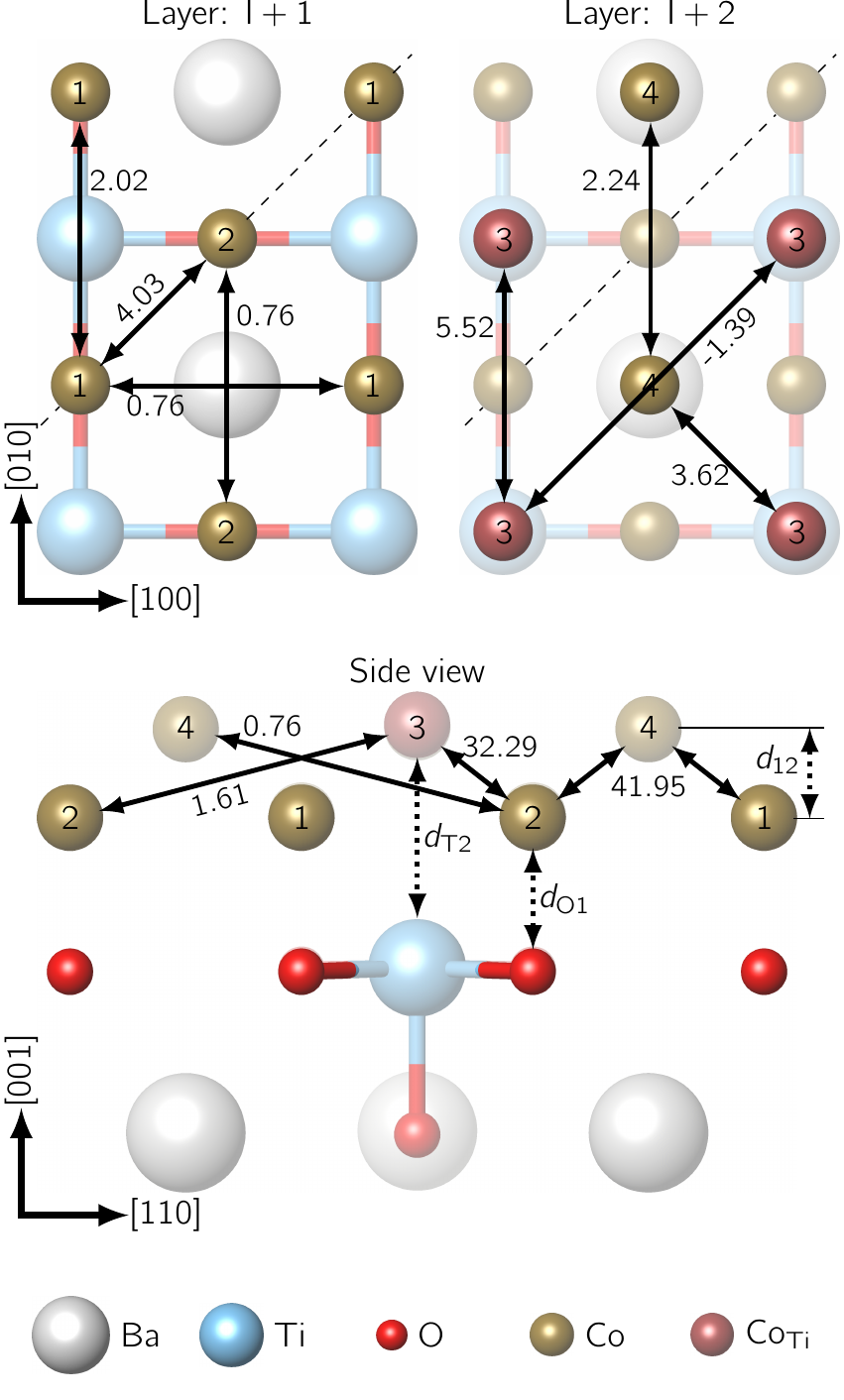}}
 \caption{Calculated exchange constants (in \si{\milli\electronvolt})
   for 2~ML Co/BTO(001) and $P_\uparrow$.  The notation is the same as
   in Fig.~\ref{fig:J1} and the numbers of the Co atoms correspond to
   those used in Fig.~\ref{fig_str}(b).  Intralayer interactions in
   the $\mathrm{I}+1$ and $\mathrm{I}+2$ layers are separated from the
   interlayer interactions shown in the side view.  The dashed lines
   in the two upper panels indicate the the plane which is shown in
   the side view. For symmetric interactions only one number is
   given.}
\label{fig:J2}
\end{figure}

Besides, we also observe induced magnetic moments at the Ti and O
atoms at the interface (see Tab.~\ref{tab_ms}). For the Ti atoms,
these induced magnetic moments are antiparallel oriented to the
direction of the magnetic moments of the Co atoms and are originated,
similar to the case of Fe/BTO(001), from the hybridization of Ti and
Co $3d$ states (see Fig.~\ref{fig_dos3d}). The strong interaction
between the host and the Co films widens the DOS of the layers at the
interface, while the $3d$ states of Co become very narrow towards the
surface, which results from the reduced coordination number of Co
atoms at the surface (in comparison to Co bulk). The impact of the
crystalline environment is very strong on the DOS and magnetic moments
of the Co atoms. As the result, the magnetic moments on Co$_\mathrm{Ti}$
are smaller by \SIrange{0.08}{0.15}{\muB} then on
Co$_\mathrm{Ba}$($\mathrm{I}+2$) due to the strong interaction between the
Co$_\mathrm{Ti}$ and the Ti states. A short distance between
Co($\mathrm{I}+3$) and Co($\mathrm{I}+2$) and a strong Co-Co hybridization
in the case of $L=3$ leads to a substantial reduction of magnetic
moments of the surface Co layer.

We have calculated the orbital moment of Co atoms along
$\vec{M}\parallel[001]$ (see Tab.~\ref{tab_mo}) and along
$\vec{M}\parallel[100]$ (see Tab.~\ref{tab_mo1}). In the last case,
the two Co atoms in each layer are nonequivalent. In the surface layer
the atoms show a strong dependence on the electric polarization of the
substrate.  For $\vec{M}\parallel[001]$ the influence of the polarization
is weak and is as large as for the Fe/BTO system.~\cite{Borek2012}

\begin{figure*}
  \center{\includegraphics[width=0.7\textwidth]{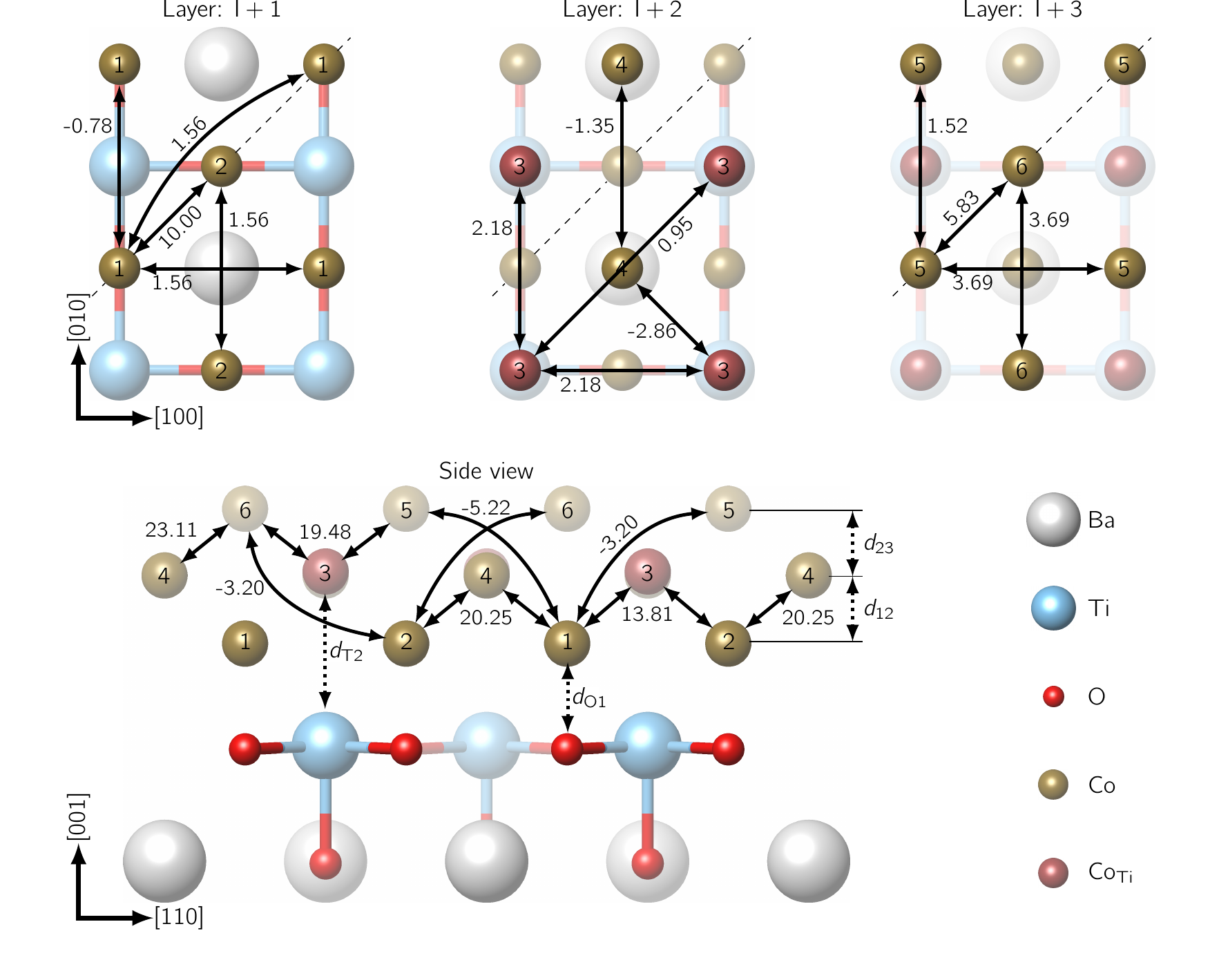}}
  \caption{Calculated exchange constants (in \si{\milli\electronvolt})
    for 3~ML Co/BTO(001) and $P_\uparrow$.  The notation is the same
    as in Fig.~\ref{fig:J1} and the numbers of the Co atoms correspond
    to those used in Fig.~\ref{fig_str}(b).  Intralayer interactions
    in the three layers $\mathrm{I}+1$, $\mathrm{I}+2$ and $\mathrm{I}+3$
    are separated from the interlayer interactions shown in the side
    view. The dashed lines in three upper panels indicate the the
    plane which is shown in the side view. For symmetric interactions
    only one number is written (see text).}
\label{fig:J3}
\end{figure*}

Duan \etal~\cite{Duan2008} predicted that the magnetic anisotropy
of thin magnetic films may be effected by the polarization of the
ferroelectric substrate. His calculations done for 1~ML Fe/BTO gave
differences for the orbital magnetic moment 
$\Delta m_\mathrm{o}=m_\mathrm{o}[001]-m_\mathrm{o}[100]$ of
\SI{0.035}{\muB} and \SI{0.021}{\muB} for $P_\uparrow$ and
$P_\downarrow$, respectively. In our former work of 1~ML Fe/BTO, we
found also an averaged anisotropy of the orbital magnetic moment
$\Delta m_\mathrm{o}$ of \SI{0.04}{\muB} and \SI{0.06}{\muB} (note
that there are two Fe atoms in the unit cell of
1~ML~Fe).~\cite{Borek2012} In case of 1~ML Co/BTO, we get the averaged
$\Delta m_\mathrm{o}=\SI{0.04}{\muB}$ for the polarization $P_\uparrow$,
which is similar to 1~ML Fe/BTO. There was no strong dependence on the
polarization on the BTO substrate in our calculations.
The anisotropies for the orbital moment will influence the preference
of the easy axis of magnetization as discussed in the next section.

On the other hand, for thicker Co films, the electronic structure can be strongly
affected by a Schottky barrier, which develops at a
semiconductor-metal interface and can lead to a spurious charge
transfer across the interface.~\cite{Stengel2011} However, in very
thin metallic films on a semiconductor, the Schottky barrier is not
significant and can not substantially modify their electronic and
magnetic structure.
  
\subsection{Magnetic interactions and  Curie temperatures}

To study magnetic interactions in ultrathin Co films on BTO(001), we
computed the exchange constants which enter the Heisenberg
model~(\ref{eq2}).  The most significant magnetic exchange
interactions for Co/BTO(001) are shown in Figs.~\ref{fig:J1} to
\ref{fig:J3} for $L=1,$ 2, 3~ML, respectively, while for the sake of
completeness all calculated coupling constants are presented in the
supplementary material.  They are always separated in interlayer and
intralayer contributions.

The key feature of the elaborated results is a very strong magnetic
coupling between the nearest neighbors. This means for the 1~ML Co
film the nearest neighbor coupling is in-plane (nearest neighbor Co
distance is \SI{2.79}{\angstrom}) with about
\SI{21}{\milli\electronvolt} (see Fig.~\ref{fig:J1}). In this case
the electronic density is mainly distributed within the $xy$-plane and
the exchange integrals represent an overlap between the electronic
wave functions.  Hence, comparing with Co bulk (hcp) the magnetic
coupling in 1~ML remains stronger despite the larger nearest neighbor
distance (in Co bulk (hcp) the nearest neighbor magnetic coupling is
$\approx\SI{13}{\milli\electronvolt}$).

In contrast, for the films with an out-of-plane component (2~ML,
  3~ML), the distances towards the Co atoms in the next layer are
shorter (\SI{2.23}{\angstrom}). Therefore, the nearest neighbor
coupling is now between those layers and the wave function overlap is
even stronger.  It leads to a strong magnetic interaction between
\SIrange{32}{42}{\milli\electronvolt} for 2~ML (see
Fig.~\ref{fig:J2}).  In the case of 3~ML, the electronic density
redistributes in all directions and, therefore, the strength of the
interlayer exchange interactions with respect to those of 2~ML is
reduced to \SIrange{13}{20}{\milli\electronvolt} (see
Fig.~\ref{fig:J3}).  Comparing these results with the exchange
constants in Co bulk, we found in general a strong asymmetry
between in-plane and out-of-plane components.  The latter are
  larger while the in-plane directions are strongly reduced (see
Figs.~\ref{fig:J2} and \ref{fig:J3}).

Furthermore, we obtain a partial mediation of the magnetic coupling by
the BTO host.  The exchange interaction between e.g. the second
nearest neighbor Co atoms depends on the underlying atom, either
$\approx\SI{-3.3}{\milli\electronvolt}$ or
$\approx\SI{0.4}{\milli\electronvolt}$ with mediating Ba or Ti,
respectively (see Fig.~\ref{fig:J1}). This fact can also be observed
for 2 and 3~ML (see Figs.~\ref{fig:J2} and \ref{fig:J3}) and is
evident from the comparison of the calculated $J_{ij}$ for supported
and unsupported thin Co films, shown in the supplementary material.
While using the same geometry for the Co films, we removed the BTO
substrate in the calculations.  Immediately, the symmetry of $x$ and
$y$ directions returns and some exchange interaction values differ
more than \SI{10}{\milli\electronvolt}.

\begin{table}
  \caption{Magnetic anisotropy energy and critical temperature for (Co)$_L$/BTO(001) 
    in dependence on the Co film thickness $L$ and the direction of ferroelectric polarization $\vec{P}$.
    $T_\mathrm{C}$ in brackets is for $\MAE=\SI{0}{\milli\electronvolt}$.
  } 
  \label{tab_MAE}
  \begin{ruledtabular}
    \begin{tabular*}{\columnwidth}{@{\extracolsep{\fill}}l D{.}{.}{-1} D{.}{.}{-1} c c}
	& \multicolumn{2}{c}{$\MAE$ (\si{\milli\electronvolt})} & \multicolumn{2}{c}{$T_\mathrm{C}$(\si{\kelvin})} \\
      $L$& \Pcoldown & \Pcolup & \Pcoldown & \Pcolup \\
      \hline
      $1$  & -1.847 & -1.483 & \multicolumn{1}{c}{360  (307)} & \multicolumn{1}{c}{298  (250)}\\ 
      $2$  &  0.680 &  0.958 & \multicolumn{1}{c}{844  (820)} & \multicolumn{1}{c}{818  (780)}\\ 
      $3$  & -0.270 & -0.438 & \multicolumn{1}{c}{567  (550)} & \multicolumn{1}{c}{580  (550)}\\ 
    \end{tabular*}
  \end{ruledtabular}
\end{table}

For the Monte Carlo simulations with the classical Heisenberg
Hamiltonian \eqref{eq2}, we have also computed the magneto-crystalline
anisotropy energy~(MAE) $\MAE=E_{[001]}-E_{[100]}$ for various Co
thicknesses and polarizations (see Table \ref{tab_MAE}).  We found an
out-of-plane magnetization direction for the cases of 1 and
3~ML~Co/BTO(001), while in the case of 2~ML~Co/BTO(001), the
magnetization direction is in-plane. The change in the MAE with the
polarization switch can reach up to \SIrange{25}{30}{\percent}. It is
remarkable, that in the 2~ML Co case, the magnetization direction is
lying within the $xy$-plane, which can be explained by the small
interlayer distance (see Table I).

Together with the calculated $J_{ij}$ parameters, those MAE values were used 
to determine the critical temperatures $T_\mathrm{C}$ in dependence on
the layer thickness $L$ and polarization $\vec{P}$ (see Table \ref{tab_MAE}). 
In general, the thin (Co)$_L$/BTO films are ferromagnetic at room temperature
for $L\geq 2$.

The $T_\mathrm{C}$ increases not monotonically with the thickness, as
expected from the large change of the magnetic coupling
parameters. The main contribution originates from the strong
interlayer coupling.  The different polarization directions have only
a small influence to the the value of $T_\mathrm{C}$ which follows
from the similar $J_{ij}$ parameters for $P_\uparrow$ or
$P_\downarrow$ (for all values see the supplementary material). As
expected, a value for MAE $|\MAE|>0$ increases the Curie
temperature (see Table \ref{tab_MAE}).

\begin{figure}
  \center{\includegraphics[width=0.70\columnwidth]{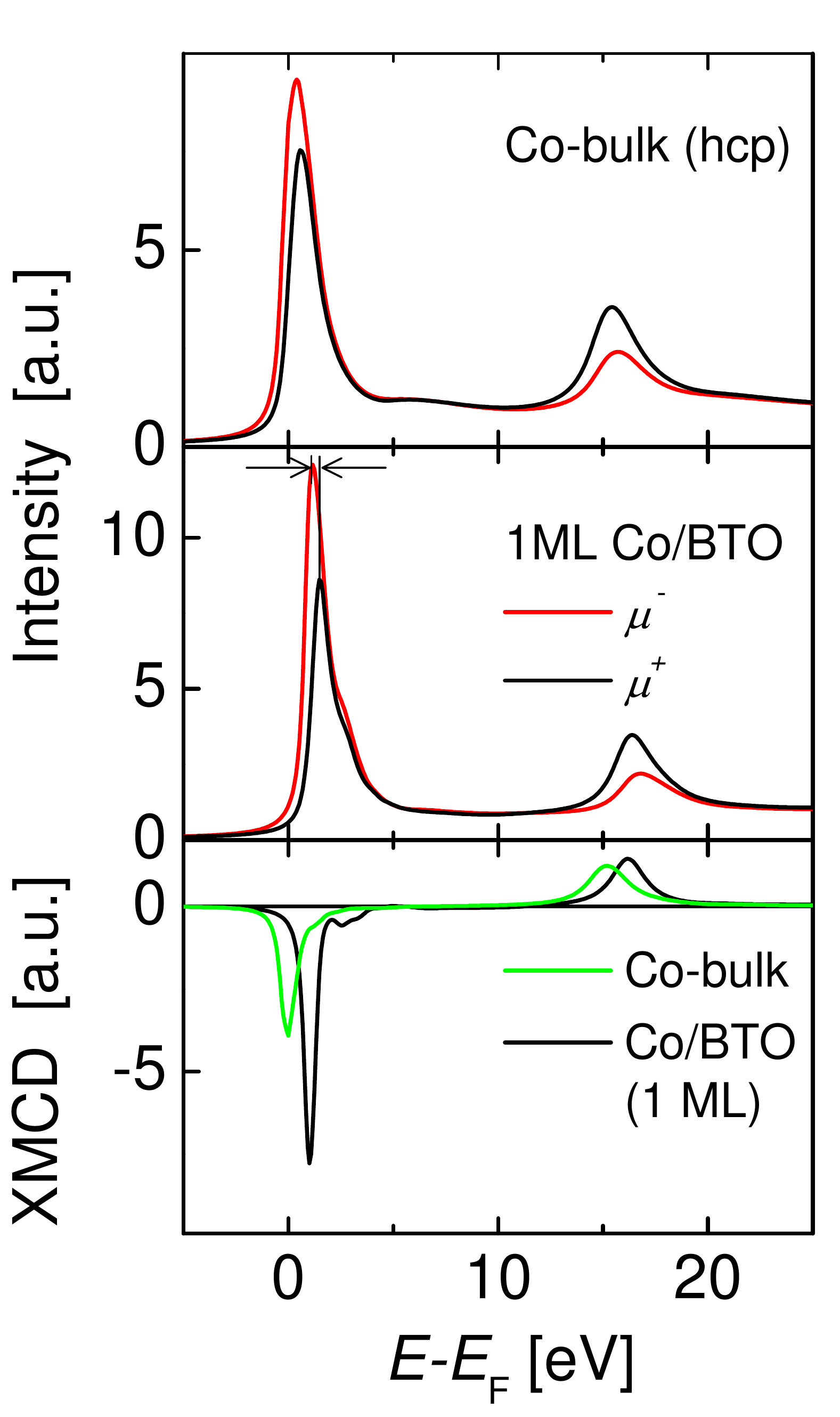}}
  \caption{ Calculated XAS (upper and middle part) and XMCD
    (lower part) of 1~ML~Co on BTO(001) ($P_\uparrow$)
    in comparison to Co bulk (hcp). Open arrows indicate
    difference in peak position (see text).}
\label{fig_xas}
\end{figure}

\begin{figure}
  \center{\includegraphics[width=0.7\columnwidth]{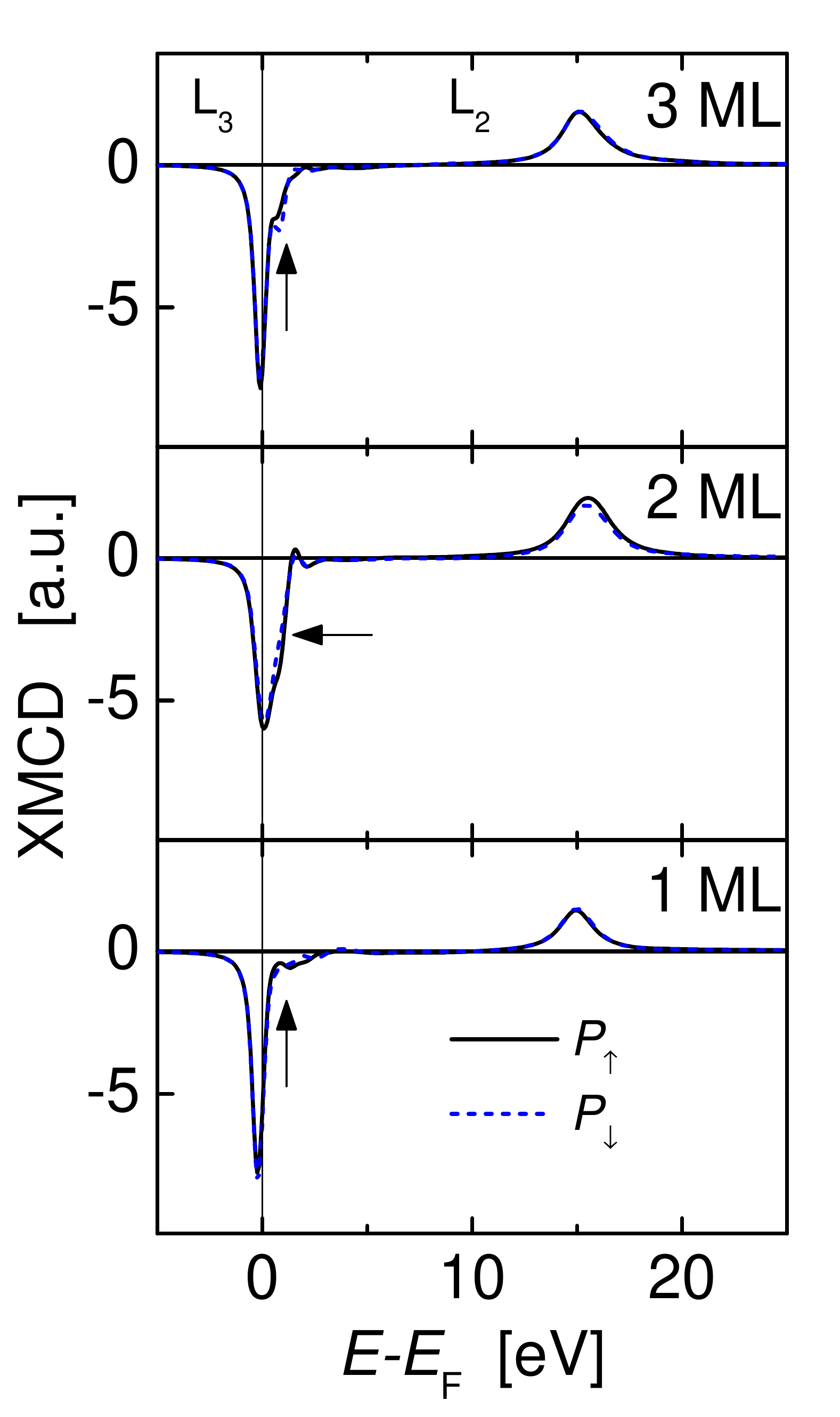}}
  \caption{Calculated XMCD  of Co L$_{2,3}$ edges in Co/BTO in dependence on 
    film thickness and electric polarization $\vec{P}$ of BTO ($P_\uparrow$ solid lines, $P_{\downarrow}$ dotted lines).
    Arrows indicate different small features (see text). 
    The thin vertical line visualizes the difference in the peak positions. }
\label{fig_xmcd}
\end{figure}

\section{XAS and XMCD}
\label{xmcd}

The Co L$_{2,3}$ edges are in the focus of our XMCD discussion.  We
have calculated x-ray absorption spectra  of right- and
left-circular polarized x-ray radiation in dependence on the Co film
thickness $L$ and the electric polarization direction $\vec{P}$ of BTO.  The
difference of the absorption coefficients 
$\Delta\mu =\mu^{+}(E)-\mu^{-}(E)\equiv\mathrm{XMCD}$ 
is normalized to the number of Co atoms in the multilayers.  
All x-ray absorption spectra were broadened by Lorentz
convolution with a core hole width of \SI{0.9}{\electronvolt} and \SI{0.25}{\electronvolt} for the L$_2$
and L$_3$ edge, respectively.  In our simulations of XAS and XMCD the
magnetization $\vec{M}$, the electric polarization $\vec{P}$ and the
incidence of light are parallel to the $z$-axis (surface normal) of
the Co/BTO(001) system.

In Fig.~\ref{fig_xas} the calculated x-ray spectra $\mu^\pm(E)$ 
(upper part) and
the related difference spectra (lower part) are shown for
1~ML Co/BTO(001) ($P_{\uparrow}$) in comparison to calculations
performed for Co bulk (hcp).  The results of our XMCD calculations
for (Co)$_L$/BTO with $L=1,2,3$ and $P_\uparrow$ and $P_\downarrow$ 
are summarized in  Fig.~\ref{fig_xmcd}.  

In all cases, we observe the well-known energy dependence of 
XMCD (Fig.~\ref{fig_xmcd}), where the intensity $\mu^+$
at the 
L$_3$ (L$_2$) edge is decreased (enhanced) due to different 
selection rules of spin-up and spin-down electrons in the 
ferromagnetic phase. The behavior is opposite at the L$_3$ (L$_2$)
edge in case of $\mu^{-}$. The dependence on the electric
polarization $\vec{P}$ of BTO is in all cases very weak
(see dotted lines in Fig.~\ref{fig_xmcd}).

In comparison to Co bulk, we found small but significant differences
concerning ultrathin films.
Note that the maxima of the L edges of left and right circularly 
polarized light are not at the same energetic position as observed for 
Co bulk XAS (see open arrows in Fig.~\ref{fig_xas}). This can be attributed 
(as in the case of Fe/BTO~\cite{Borek2012}) to the 
cancellation of degenerate $d$ states at the interface or/and surface of the 
film.

In the energy dependence of the L$_3$ edge, we observe a small
structure of the XMCD peak at higher energy 
(see closed arrows in Fig.~\ref{fig_xmcd}) in dependence on layer
thickness.  The appearance of this structure can also be explained if we
consider the spin- and layer-resolved DOS (not shown here) of $d$
states as demonstrated for the system Fe/BTO.~\cite{Borek2012}
In general, it is possible to separate all the different contributions by means of
layer-resolved XMCD calculations. This dependence on layer thickness should 
show up in corresponding experiments.

\begin{figure}
  \center{\includegraphics[width=0.7\columnwidth]{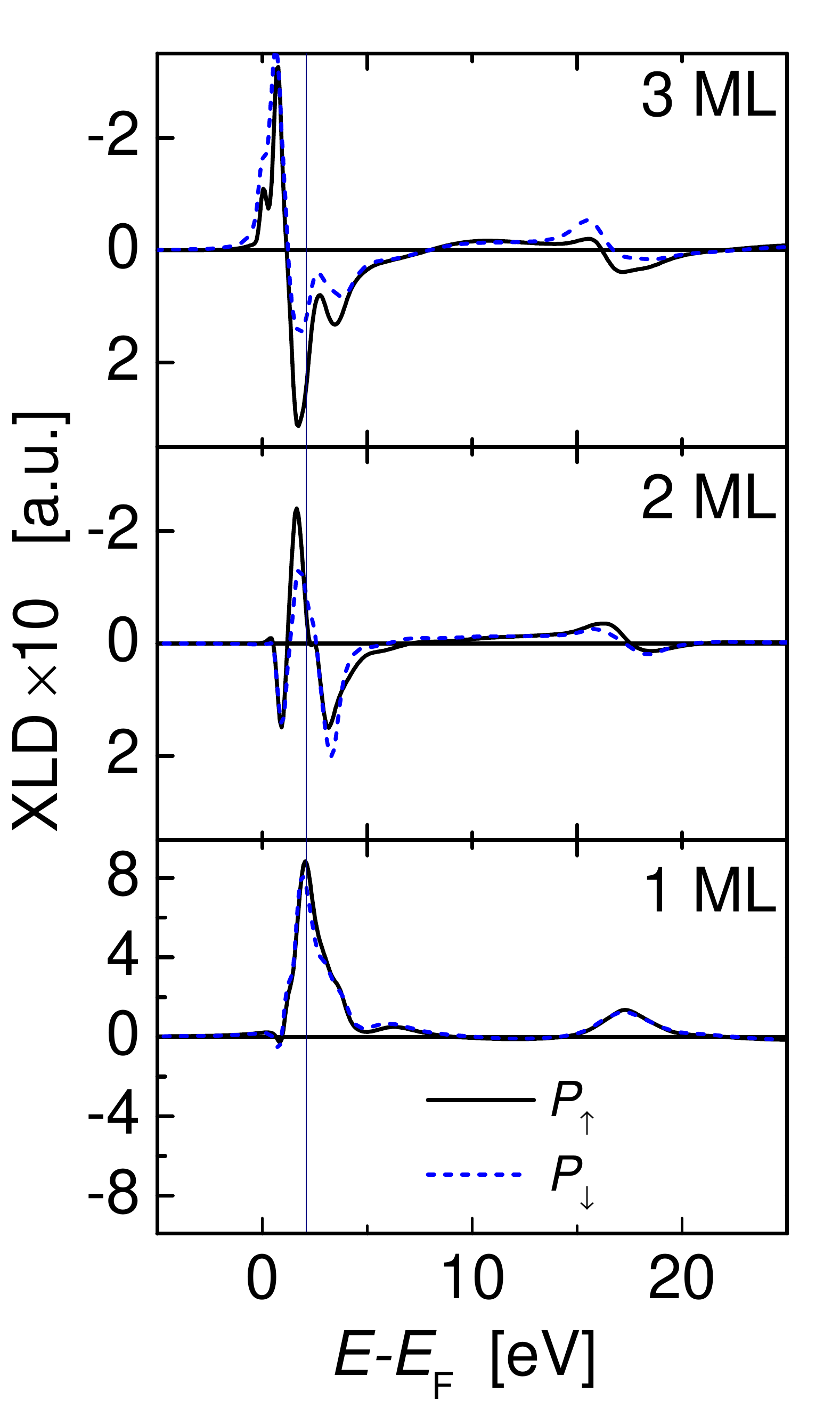}}
  \caption{Calculated XLD of Co L$_{2,3}$ edges in Co/BTO for
    polarization $\vec{P}$ ($P_\uparrow$ solid lines, $P_{\downarrow}$ dashed lines).
    The thin vertical line visualizes the difference in the peak positions. }
\label{fig_xld}
\end{figure}

\section{X(M)LD}
\label{xmld}

X-ray absorption linear dichroism can be applied in different modes.~\cite{Stoehr2006}
In nonmagnetic materials with cubic symmetry the x-ray absorption intensity 
is independent on the orientation of the polarization vector of light ($\vec{u}$)
relative to the sample. As the symmetry is lowered (like in a film), the intensity
is directly proportional to the number of empty valence states in the
direction of $\vec{u}$. The polarization vector $\vec{u}$ acts like a search
light for the direction of maximum and minimum empty valence states.~\cite{Stoehr2006}
In case of 
a magnetic sample with cubic symmetry, the spin-orbit coupling leads to magnetic
linear dichroism, where the x-ray absorption intensity is different for $\vec{u}$
aligned parallel and perpendicular to the spin direction.~\cite{Stoehr1998} 
This kind of dichroism is often considered in the discussion 
of magneto crystalline effects (for example in Ref. \onlinecite{Kunes2003}).

In our magnetic ultrathin 
Co films on BTO we have both natural and magnetic linear dichroism effects.
First, we present results for a fixed direction of magnetization
$\vec{M}\parallel[001]$ and linearly polarized light with polarization
along $[001]$ ($\mu^z$) and $[100]$ ($\mu^x$), respectively.  The
difference $\mathrm{XLD}=\mu^{z}(E)-\mu^{x}(E)$ between  both x-ray absorption
spectra is shown in Fig.~\ref{fig_xld}.

The strong dependence of the XLD on film thickness can be related to the 
occupation of $d$ orbitals.
A detailed
analysis is possible concerning the single $(p_j,m_j)$ contributions of the
initial $p_{3/2}$ ($p_{1/2}$) state to the $p\rightarrow d$ transitions
of the different Co absorbers in the film.  In the 1~ML case, we get the main
positive contribution from the transition of $m_j=-1/2$ into
$d_{3z^2-r^2}$ states. The contributions of $d_{xz}$ and $d_{yz}$ are
small because of the strong interaction between Co and Ti atoms at the
interface. The XLD of 2~ML reflects the change of occupancy between
in-plane and out-of-plane orbitals.  In the 3~ML case, we found that
the behavior is similar to the Co bulk, but with much more fine
structures.  These fine structures are related to the lifting of
degeneracy in the final $d$ state in the thin film.

Besides, we found that the XLD is much more sensitive concerning the
dependence on the polarization of the BTO substrate as the XMCD. 
This emphasize how the reversal of 
polarization $\vec{P}$ changes the occupation of $d$ states.

We have also investigated x-ray absorption in dependence on the magnetization direction
$\vec{M}$ for fixed polarization $\vec{u}\parallel[001]$ but spin-orbit interaction and
crystal field are both to weak to eliminate magneto crystalline anisotropy. 

\section{Conclusion}
\label{concl}
In this first-principles study of structural,
electronic, and magnetic properties of ultrathin Co layers on the
BaTiO$_3$(001) substrate, we showed, that the crystalline
structure of (Co)$_L$/BTO(001) ($L=1,2,3$) interfaces is very similar to
the one of (Fe)$_L$/BTO(001) films presented in our previous
study.~\cite{Fechner2008} Additionally, we found also for the Co layers 
only a small dependence on the polarization direction of the 
substrate for all investigated properties. 
However, in contrary to the Fe/BTO(001)
case, Co films on BTO(001) exhibit a stable ferromagnetic ordering at
room temperature which depends strongly on the layer thickness and is 
the strongest for $L=2$. The
magnetic interaction is mainly featured by the coupling between the
nearest neighbors. For $L\ge 2$ cases, the strongest interactions arise
between the adjacent layers, while the intralayer magnetic coupling
was found to be rather weak. This results from short
interlayer distances leading to a strong hybridization between Co $3d$
electrons of adjacent layers. Surprisingly, the easy axis turns for 2~ML
from a out-of-plane magnetization (in $L=1$ and 3) to a in-plane magnetization, 
which was obtained 
from the direct calculation of the MAE or the calculated orbital moments.

For a comparison to possible experimental measurements,
we simulated x-ray absorption spectra and related x-ray magnetic
circular and (magnetic) linear dichroism to trace
the change of the spectra under the polarization switching in the BTO
substrate. While the XMCD depends only weakly on the
substrate polarization, similar to our previous study for
(Fe)$_L$/BTO(001) films,~\cite{Borek2012} the XLD shows indeed significant changes
under polarization switching, which can probably be observed experimentally.
\begin{acknowledgments}
  We are grateful for financial support by Deutsche
  Forschungsgemeinschaft in the framework of SFB762 "Functionality of
  Oxide Interfaces".  The authors would like to thank Jan Min\'ar and
  Hubert Ebert for deployment and support of the SPR-KKR program and
  helpful discussions.
\end{acknowledgments}

\bibliography{jprb,lib}

\end{document}